\documentclass[prl,reprint,amssymb,twocolumn,noeprint,superscriptaddress,longbibliography]{revtex4-2}

 \usepackage[utf8]{inputenc}
\usepackage{amsmath}
\usepackage{graphicx} 
\usepackage{physics}
\usepackage[usenames,dvipsnames]{color}
\usepackage{amsfonts}
\usepackage{bm}
\usepackage{amssymb}
\usepackage{braket}
\usepackage{multirow}
\usepackage{epstopdf}
\usepackage[normalem]{ulem}

\usepackage[T1]{fontenc}

\usepackage[caption=false]{subfig}

\usepackage{hyperref}
\hypersetup{
colorlinks=true,
linkcolor=blue,
citecolor=blue,
filecolor=green,
urlcolor=blue,
}

\begin{document}

\title{Rotational decoherence dynamics in ultracold molecules induced by a tunable spin environment: The Central Rotor Model}

\author{Timur V. Tscherbul} 
\affiliation{Department of Physics, University of Nevada, Reno, Nevada, 89557, USA}
\author{Lincoln D. Carr}
\affiliation{Quantum Engineering Program and Department of Physics, Colorado School of Mines, Golden, Colorado, 80401, USA}
\date{\today}

\begin{abstract} 
We show that quantum rotational wavepacket dynamics in molecules can be described by a new system-environment model, 
 which consists of a rotational subsystem coupled to a magnetically tunable spin bath  formed by the nuclear spins within the molecule. 
The  central rotor model shares similarities with the paradigmatic central spin model, but features much richer rotational dynamics that is sensitive to 
the molecule's environment, which can be initiated and probed with short laser pulses used to control molecular orientation and alignment. 
We present numerical simulations of the nuclear-spin-bath-induced rotational decoherence dynamics of  KRb molecules, which  exhibit remarkable sensitivity to an external magnetic field. Our results
 show that ultracold molecular gases provide a natural platform for the experimental realization of the CRM.
\end{abstract}
\maketitle

Achieving robust quantum control over rotational motion is essential in a wide variety of research fields, including molecular reaction dynamics~\cite{Zhang:16,Zhao:17,Pan:20}, orientation and alignment~\cite{Cai:01,Stapelfeldt:03,Koch:19}, ultracold chemistry~\cite{Krems:08,Balakrishnan:16,Klein:17,Koch:19,Bohn:17}, and quantum science \cite{Cornish:24,DeMille:02,Yelin:06,Gorshkov:11,Albert:20,Asnaashari:23}.
Aided by recent advances in molecular cooling and trapping~\cite{DeMille:10,Hummon:13,Barry:14,Truppe:17,Anderegg:18,Mitra:20,VazquezCarson:22,Burau:23,Hallas:23,Anderegg:23,Vilas:24,Son:20,Kaufman:21}, high-fidelity quantum control of molecular rotations over second-long timescales has been demonstrated experimentally \cite{Gregory:24,Cornish:24}.
However, interactions with an uncontrolled environment limit the fidelity of quantum control. Examples include noisy electromagnetic fields due to the trapping light~\cite{Kotochigova:10,Neyenhuis:12,Sesselberg:18}, inhomogeneous external magnetic fields~\cite{Sawant:20}, and blackbody radiation~\cite{Hoekstra:07,Haas:19,Vilas:23}.
Understanding and mitigating these environmental interactions requires the development of rigorous system-bath models, in which molecular rotational degrees of freedom interact with a larger environment ~\cite{SchlosshauerBook,BreuerBook}. 
Such models have attracted much attention in the context of  decoherence of rotational wavepackets in the gas phase (via collisions)~\cite{Ramakrishna:05,Ramakrishna:06,Milner:14,Milner:14b,Khodorkovsky:15,Zhang:19,Ma:19,Bournazel:23,Wang:24,Stickler:18a,Stickler:18b} and in bosonic environments, such as superfluid helium nanodroplets~\cite{Toennies:04,Pentlehner:13,Chatterley:19,Schmidt:15,Schmidt:16,Lemeshko:17,Shepperson:17,Qiang:22,Milner:23,Milner:24}. Extensive experimental studies of rotational wavepacket dynamics have been carried out~\cite{Stapelfeldt:03}
yielding invaluable information about molecular  orientation and alignment in various environments. Theoretical studies of this dynamics have also been reported, most recently in the context of the angulon quasiparticle, which describes a quantum rotor dressed by many-body excitations of a bosonic environment~\cite{Schmidt:15,Schmidt:16,Lemeshko:17,Shepperson:17,Qiang:22}. 

A much less well-explored flavor of quantum dynamics arises due to the interaction of molecular rotation {\it with a spin environment}.
The vast majority of molecular species possess magnetic nuclei, which  interact with  rotational motion via hyperfine interactions~\cite{Brown:03,Hirota:85,Zare:88,Yachmenev:17}, leading to nuclear spin depolarization~\cite{Altkorn:85,Cool:95,Rutkowski:04,Bartlett:09} and loss of alignment~\cite{Thomas:18,Yachmenev:19,Thesing:20}. These phenomena are of both fundamental and practical significance because rotational wavepackets can serve as a powerful probe of intra- and intermolecular interactions~\cite{Ramakrishna:05,Ramakrishna:06,Lu:24} and many-body host environments~\cite{Schmidt:15,Schmidt:16,Lemeshko:17,Shepperson:17,Qiang:22}, and can be used to create long-lasting molecular spin polarization~\cite{Rakitzis:05}. 
 The effects of hyperfine interactions on molecular alignment were studied theoretically~\cite{Rakitzis:05,Grohmann:11,Floss:12,Thesing:20} and recently observed experimentally for I$_2$ molecules~\cite{Thomas:18}.
However, these studies have focused on short timescales and  cold (a few Kelvin) conditions found in supersonic expansions, resulting in a largely uncontrolled spin environment. Recent advances in the production of ultracold ($\simeq$~1~$\mu$K) diatomic~\cite{Bohn:17} and polyatomic~\cite{Mitra:20,Langen:24} molecules have demonstrated   high-fidelity quantum control over both rotational and nuclear spin degrees of freedom, suggesting the possibility of  studying rotational wavepacket dynamics under the influence of a tunable spin environment. In addition to providing  novel insights into rotational dynamics in hitherto unexplored regimes, these capabilities open the door to exploring the dynamics of quantum information, thermalization and entanglement between the different molecular degrees of  freedom (such as rotations and spins) as recently demonstrated for isolated quantum many-body systems \cite{LewisSwan:19}.

Here, we show that molecular rotational wavepacket dynamics can be rigorously described by a novel system-environment model, in which a central subsystem  composed of a quantum rotor (or, more generally, an asymmetric top)  is coupled to a spin bath, as illustrated in Fig.~1(a): the \emph{Central Rotor Model} (CRM).  
Here, the  subsystem's degrees of freedom are those of the molecule's mechanical rotation, and the spin bath is formed by the nuclear spins within the molecule. The system-bath coupling is provided by the hyperfine spin-rotation interactions (see below).
The CRM is analogous to the central spin model  (CSM)~\cite{Prokofev:00,SchlosshauerBook}, which describes the dynamics of a single electron spin interacting with a localized (nuclear) spin bath, and features highly coherent dynamics due to its integrability, i.e., the existence of an extensive number of integrals of motion~\cite{Ashida:19}. It serves as the archetypal model of spin dynamics in a wide range of physical systems ranging from solid-state  spin qubits such as nitrogen-vacancy centers in diamond~\cite{Childress:06,Hanson:08,Hall:14}, semiconductor quantum dots~\cite{Sousa:05,Shenvi:05,Chekhovich:13,Bechtold:15,Witzel:08,Witzel:12} and molecular magnets~\cite{Prokofev:00} to Rydberg impurities in ultracold bosonic environments~\cite{Ashida:19},  and  Rydberg atoms surrounded by polar molecules~\cite{Dobrzyniecki:23}. 
However, in contrast to the CSM, the Hilbert space of the central subsystem in the CRM is infinite-dimensional, resulting in uniquely molecular features absent in the CSM, such as rotational wavepacket dynamics~\cite{Stapelfeldt:03}.  
Using exact diagonalization (ED) simulations based on realistic  molecular Hamiltonians, we explore the dynamics of quantum information (thermalization and entanglement) in the CRM for KRb molecules, and find that it is strongly sensitive to the external magnetic field.
Our results offer a novel perspective  on rotational wavepacket dynamics in molecules based on the theory of open quantum systems, and suggest efficient computational approaches toward simulating these dynamics and extending rotational coherence times, a crucial goal for emerging applications of ultracold molecules in quantum information science~\cite{Langen:24,Cornish:24}.

As an experimental realization of the CRM,  we propose an ensemble of polar molecules trapped in optical tweezers as recently realized experimentally~\cite{Kaufman:21,Vilas:24}.
As shown in Fig.~1, the molecules are initialized in the ground rotational state, where the rotational system is fully decoupled from the nuclear spin bath. The rotational dynamics is triggered by a short off-resonant laser pulse,  leading to the formation of a rotational wavepacket~\cite{Stapelfeldt:03}.
 The wavepacket evolves in time and decoheres due to the interaction with the nuclear spin bath. Molecular rotational dynamics is probed in real time by photoionizing the molecules using femtosecond laser pulses~\cite{Schouder:22}.

 \begin{figure}[t]
 \centering
 \includegraphics[width=1.0\columnwidth, trim = 30 170 30 170]{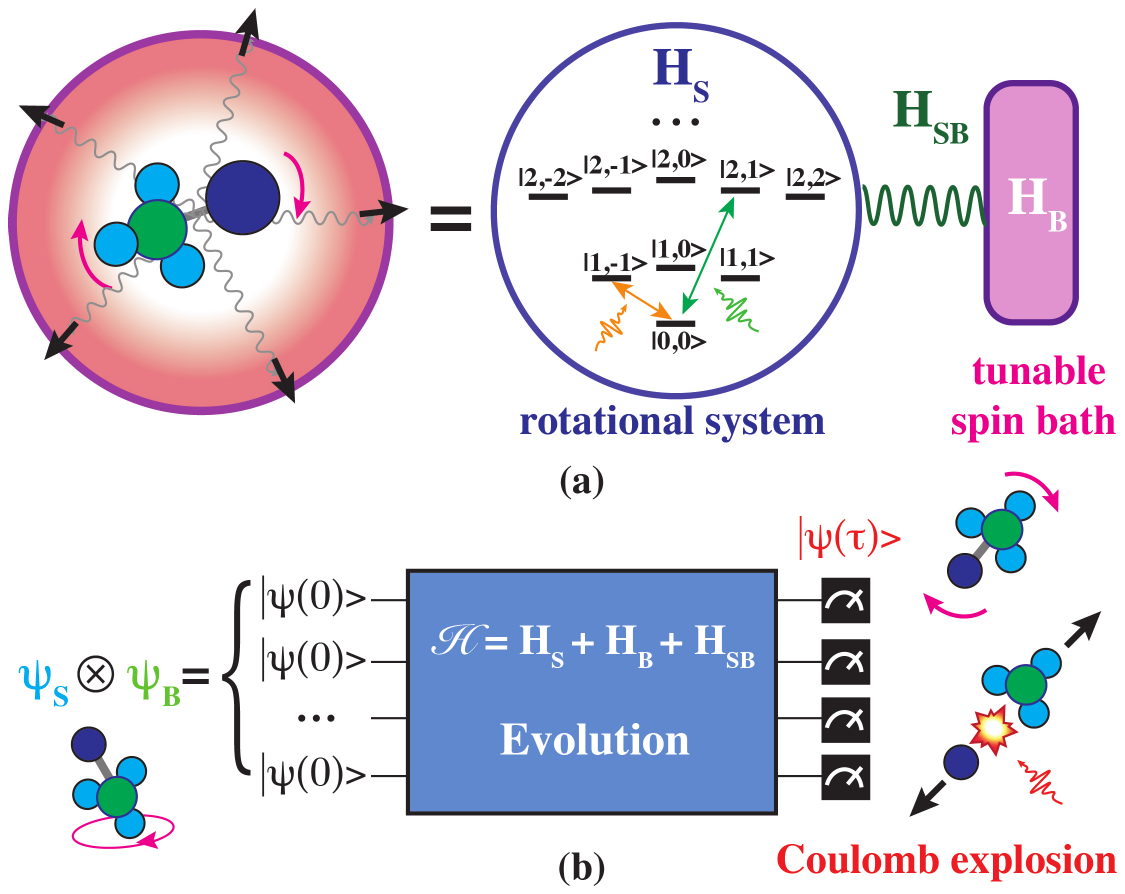}
 \caption{  (a)  Schematic of the CRM. The molecular Hamiltonian is separated into the  rotational subsystem ($\hat{H}_S$), the nuclear spin bath ($\hat{H}_B$), and the system-bath interaction ($\hat{H}_{SB}$). (b) Experimental protocol: The molecules are initialized in the ground rotational state, where the rotational and nuclear spin wavefunctions are fully decoupled.
 The time evolution of the rotational wavepacket is probed by applying a second femtosecond laser pulse, which causes a Coulomb explosion followed by the detection of ionized fragments~\cite{Schouder:22}. }
 \label{fig1cartoon}
\end{figure}

{\it Central Rotor Model}.
We begin by showing how the generic Hamiltonian of a polyatomic molecule can be mapped to that of a central quantum rotor interacting with a tunable spin bath. 
We assume that (i) the molecule resides in a specific (e.g., ground) electronic and vibrational state, whose lifetime is large compared to all other timescales of interest,
(ii)  the  electronic state  is of $^1\Sigma$ symmetry, and (iii) the polyatomic molecule has no large-amplitude vibrational modes (such as the umbrella inversion mode in NH$_3$).
These assumptions are valid for most diatomic and medium-sized polyatomic molecules~\cite{Barone:21,Bizzocchi:22}, including those cooled and trapped in recent experiments, such as ultracold alkali-dimers~\cite{Bohn:17}, and polyatomic molecules formaldehyde, CH$_3$F,  CF$_3$H, and  CF$_3$CCH~\cite{Zeppenfeld:12,Chervenkov:14}. Molecular radicals with nonzero electron spin, such as laser-coolable SrF, CaH, CaF, and YO~\cite{DeMille:10,Hummon:13,Barry:14,Truppe:17,Anderegg:18,Mitra:20,VazquezCarson:22,Burau:23,Hallas:23,Anderegg:23,Vilas:24} could be described by an extended CRM which includes the  electron spin degree of freedom~\cite{SM}.
The molecular Hamiltonian may  be written as
\cite{Wofsy:70,Chapovsky:99,Bizzocchi:22,Bowater:73}
\begin{equation}\label{Hmol}
\hat{H} = \hat{H}_\text{rot} +  \hat{H}_\text{hfs}  + \hat{H}_Z,
\end{equation}
where the asymmetric top Hamiltonian in the rigid-rotor approximation~\cite{BunkerBook,Zare:88,Green:76} 
\begin{equation}\label{Hrot}
\hat{H}_\text{rot}= A\hat{N}_x^2 + B\hat{N}_y^2 + C\hat{N}_z^2,
\end{equation}
is expressed in atomic units via the molecule-fixed principal-axes components of the rotational angular momentum operator $\hat{\mathbf{N}}$ and $A$, $B$, and $C$ are the  rotational constants~\cite{Zare:88}.
The hyperfine Hamiltonian~\cite{Wofsy:70,Chapovsky:99,Bizzocchi:22,Bowater:73}
\begin{multline}\label{Hhfs}
 \hat{H}_\text{hfs} =
  \frac{1}{2}\sum_{k,i} [T^k(\bm{C}_i),T^k(\mathbf{N},\mathbf{I}_i)]_+ +   \sum_{i<j}c_{ij} \hat{\mathbf{I}}_i \cdot \hat{\mathbf{I}}_j \\
 - \sum_{i<j} \sqrt{6}\, T^2(\bm{C}_{ij})\cdot T^2(\hat{\mathbf{I}}_i,\hat{\mathbf{I}}_j) 
  -\sum_{i}T^2(\nabla \mathbf{E}_i)\cdot T^2(\bm{Q}_i)
\end{multline}
includes the nuclear spin-rotation and spin-spin interactions described by the tensors  $\bm{C}_i$ and $\bm{C}_{ij}$. The nuclear electric quadrupole (NEQ) interaction given by the last term  in Eq.~\eqref{Hhfs} provides  the dominant  contribution to $\hat{H}_\text{hfs}$, and is expressed via the nuclear quadrupole moment and electric field gradient tensors ($\bm{Q}_i$ and  $\nabla\mathbf{E}_i$) on the $i$-th nucleus with nuclear spin $\hat{\mathbf{I}}_i$.
In Eq.~\eqref{Hhfs}, $T^k(\hat{\mathbf{O}})$ is a spherical tensor of rank $k$ composed of the components of $\hat{\mathbf{O}}$, $T^k(\hat{\mathbf{O}}_1,\hat{\mathbf{O}}_2)$ is a spherical tensor product of operators $\hat{\mathbf{O}}_1$ and $\hat{\mathbf{O}}_2$~\cite{Brown:03}, and $[,]_+$ denotes the anticommutator.
  The Zeeman interaction of the molecule with an external magnetic field of strength $B$  directed along the space-fixed quantization axis $Z$ is~\cite{Hirota:85,Bowater:73}
\begin{equation}\label{HZeem}
 \hat{H}_Z = - g_rB \mu_N \hat{N}_Z   - g_N \mu_N \hat{I}_Z B, 
\end{equation}
where $\mu_0$ and $\mu_\text{N}$ are the electron and nuclear Bohr magnetons,  and $g_S$, $g_r$, and $g_N$ are the electronic, rotational, and nuclear spin $g$-factors.
The total nuclear spin operator  $\hat{\mathbf{I}}=\sum_i \hat{\mathbf{I}}_i$.

We now make the crucial observation that   the molecular Hamiltonian~\eqref{Hmol} can be  rearranged in the system-bath form 
\begin{equation}\label{HmolSB}
\hat{H} = \hat{H}_S +  \hat{H}_{B} +  \hat{H}_{SB}\,,
\end{equation}
where the system  Hamiltonian $\hat{H}_S = \hat{H}_\text{rot} - g_rB \mu_N \hat{N}_Z$ is identified with the rotational Hamiltonian~\eqref{Hrot} plus a small rotational-Zeeman interaction.  The spin bath Hamiltonian describes molecular spins and their interactions with an external magnetic field 
\begin{equation}\label{HB}
 \hat{H}_B = 
 \sum_{i<j}c_{ij} \hat{\mathbf{I}}_i \cdot \hat{\mathbf{I}}_j 
  - g_N \mu_N \hat{I}_Z B. 
\end{equation}
Finally, the system-bath interaction $ \hat{H}_{SB}$ is given by
\begin{multline}\label{HSB}
\hat{H}_{SB}= -\sum_{i}T^2(\nabla \mathbf{E}_i)\cdot T^2(\bm{Q}_i) \\ 
 + \frac{1}{2}\sum_{k,i} [T^k(\bm{C}_i),T^k(\mathbf{N},\mathbf{I}_i)]_+ 
 - \sum_{i<j} \sqrt{6} T^2(\bm{C}_{ij})\cdot T^2(\hat{\mathbf{I}}_i,\hat{\mathbf{I}}_j)
\end{multline}
Equations~\eqref{HmolSB}-\eqref{HSB}  define the CRM,  a central result of this work. All the terms in Eq.~\eqref{HSB} depend on both the rotational and spin degrees of freedom, with the former dependence arising from the transformation of molecule-fixed tensors, such as $T^2(\nabla \mathbf{E}_i)$ and $ T^2(\bm{Q}_i)$ to the space-fixed coordinate frame~\cite{Bowater:73}, as described below.

{\it CRM dynamics with ultracold alkali-dimer molecules}. To flesh out the concept of the CRM and explore the feasibility of its experimental realization with ultracold polar molecules, 
we carry out ED calculations for the paradigmatic alkali-dimer molecule KRb~\cite{Ni:08,Ospelkaus:10prl}. Our analysis is also applicable to other alkali-dimer molecules cooled and trapped in recent experiments, such as NaK~\cite{Park:15}, NaCs~\cite{Stevenson:23}, and RbCs~\cite{Takekoshi:14,Ruttley:23}.
A possible experimental protocol for observing CRM dynamics is sketched in Fig.~1(b). An ensemble of ultracold trapped KRb molecules is  initialized in a single nuclear spin sublevel of the ground rotational state $|NM_N\rangle|M_{I_1}M_{I_2}\rangle=|00\rangle |-\!4\,\,\frac{1}{2}\rangle$, as demonstrated experimentally~\cite{Ospelkaus:10}.  Here, $|NM_N\rangle$ are the eigenstates of molecular rotational angular momentum  $\hat{\mathbf{N}}^2$ and its projection of the quantization axis $\hat{N}_Z$, and $|M_{I_1}M_{I_2}\rangle=|I_1M_{I_1}\rangle |I_2M_{I_2}\rangle$ are the eigenstates of the nuclear spin operators $\hat{\mathbf{I}}_i^2$ and $\hat{\mathbf{I}}_{i_Z}$ (the index $i=1,2$ labels the nuclei).
 A weak  dc magnetic field is applied to decouple the rotational and spin degrees of freedom, resulting in the product initial state of the form $|\psi(t=0)\rangle =  |00\rangle \otimes |\psi_s\rangle$, where $|\psi_s\rangle$ is a nuclear spin function~\cite{SM}.
At time $t=0$, an off-resonant ``kick'' pulse is applied to the $N=0$ molecules  creating a spin-rotational wavepacket 
$|\psi(t=0^+)\rangle =   |\psi_R(0^+)\rangle \otimes \rangle |\psi_s\rangle$,
where $ |\psi_R(0^+)\rangle=  \sum_{N} d_{N0} |N0\rangle   $ is the rotational part of the wavepacket ~\cite{Stapelfeldt:03}, 
and $d_{N0}$ are complex coefficients which depend on the dimensionless power parameter $P$ that characterizes the pulse strength~\cite{Leibscher:03,Leibscher:04}.
We assume that the pulse is linearly polarized $(M_N=0)$ and fast compared to the rotational period $t_R=1/(2 B_e)$, which will serve as a natural unit of time.

Upon excitation, the wavepacket evolves under the alkali-dimer Hamiltonian
 which is a particular case of  Eq.~\eqref{HmolSB} for diatomic molecules~\cite{Brown:03,Aldegunde:08}. We will consider the $^{40}$K$^{87}$Rb molecule with two magnetic nuclei, $I_1=4$ ($^{40}$K) and $I_2=3/2$ ($^{87}$Rb). Accordingly, we identify the system Hamiltonian with $\hat{H}_S = B_e\hat{N}^2$, the spin bath Hamiltonian  with $\hat{H}_B = c_{4} \hat{\mathbf{I}}_1 \cdot \hat{\mathbf{I}}_2   - g_N \mu_N (\hat{I}_{Z_1} + \hat{I}_{Z_2}) B$, and the system-bath coupling with~\cite{Gorshkov:11}
\begin{multline}
\hat{H}_{SB} = \sum_{m,p} (-1)^p C_p^2(\theta,\phi) \frac{\sqrt{6} (eqQ)_m}{4 I_m(2I_m-1)}
 T^2_{-p}(\hat{\textbf{I}}_m,\hat{\textbf{I}}_m) \\
 + \sum_{i=1,2}  c_i \hat{\mathbf{N}}\cdot \hat{\mathbf{I}}_i - c_3 \sqrt{6}\sum_p (-1)^p C^2_{-p}(\theta,\phi) T^2_p(\hat{\mathbf{I}}_1,\hat{\mathbf{I}}_2),
\end{multline}
where the dominant NEQ interaction is expressed  via  the coupling constants $(eqQ)_m$ and renormalized spherical harmonics $C^k_p(\theta,\phi)=\sqrt{{4\pi}/(2k+1)} Y_{kp}(\theta,\phi)$, and  the spherical polar angles $\theta$ and $\phi$ specify the position of the molecular axis with respect to $Z$~\cite{Gorshkov:11,Hermsmeier:24}.
Here, we are interested in rotational dynamics, so it is convenient to define the reduced density operator of the rotational subsystem by tracing out the nuclear spin degrees of freedom~\cite{SM} 
$\hat{\rho}_R = \text{Tr}_{\!B} \hat{\rho} = \sum_{M_{I_1},M_{I_2}} \langle M_{I_1}M_{I_2} | \hat{\rho} | M_{I_1} M_{I_2}\rangle$, where
$\hat{\rho}(t)=|\psi(t)\rangle\langle \psi(t)|$ is the full density matrix. This gives the  rotational density matrix $\hat{\rho}_R(t)$, from which all rotational observables of interest can be computed. Of particular importance  is the expectation value of the alignment cosine, $\langle \cos^2\theta\rangle = \text{Tr} \left[\hat{\rho}_R \cos^2\theta \right]$ 
which can be measured experimentally~\cite{Stapelfeldt:03}.

 \begin{figure}[t]
 \centering
 \includegraphics[width=0.8\columnwidth, trim = 20 0 0 -10]{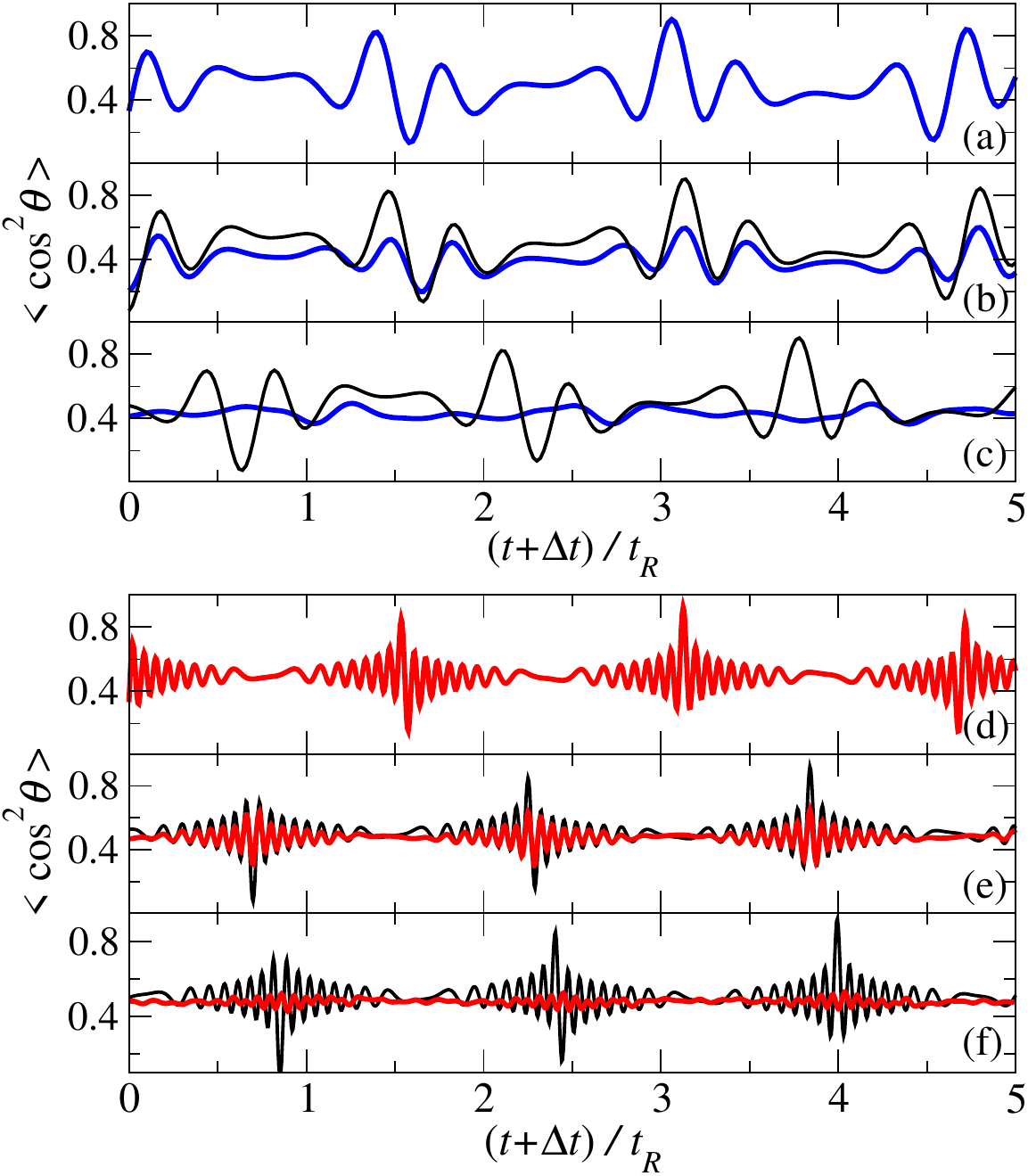}
 \caption{Time evolution of rotationally warm [(a)-(c), $P=10$]  and rotationally hot [(d)-(f), $P=50$] KRb wavepackets at $B=10$~G as measured by the expectation value of the orientation cosine $\langle \cos^2\theta\rangle(t)$. Quantum dynamics via ED (blue and red lines), purely rotational dynamics (black lines).   A kick pulse of strength $P$ and duration $t_p\ll t_R$ is applied to molecules at $t=0$, and the dynamics is plotted after the delay times (in units of $t_R=1/2 B_e$): (a) $\Delta t=0$, (b) $2\times 10^5$, (c) $2\times 10^6$. Panels (d)-(f):  Same as in panels (a)-(c) but for a stronger kick pulse with $P=50$ at (e) $\Delta t=0$, (e) $2\times 10^6$, and  (f) $2\times 10^7$.}
 \label{fig1cartoon}
\end{figure}

Figure 2 shows the time evolution of the alignment cosine calculated for rotationally ``warm'' vs. rotationally ``hot'' wavepackets 
containing $\langle {N}\rangle\simeq 7$ and 36 rotational states, respectively.
At short times, both types of rotational wavepackets exhibit periodic oscillations and revivals, which are well understood theoretically and characterized experimentally~\cite{Seideman:99,Stapelfeldt:03}. 
These structures are governed by the rotational kinetic energy, and  spin degrees of freedom play no role: the reduced dynamics shown  in Figs.~2(a) and (d) is identical to purely rotational dynamics $\hat{\rho}_R^{(0)}(t)=|\psi_R(t)\rangle \langle \psi_R(t)|$ calculated for a rigid rotor with no coupling to the nuclear spins.

The situation changes dramatically at later times. As shown in Figs. 2(b)-(c) and (e)-(f), the nuclear spin bath starts to alter the rotational dynamics after $\sim$10$^5$ rotational periods, reducing the amplitude of the coherent oscillations, and causing increasingly large deviations from purely rotational evolution. 
The decoherence timescale observed is consistent with the magnitude of the NEQ interaction, which provides the dominant contribution to the system-bath coupling in KRb. Because it is much smaller than the rotational constant ($|eQq_2|/B_e\simeq 0.8\times 10^{-3}$~\cite{Aldegunde:08}), the effects of the nuclear spin bath are weak at short times, as observed in Figs.~2(a) and 2(d).
The bath has a weaker effect on the rotationally hot wavepacket due to its larger average rotational energy. In the long-time regime ($t/t_R \gg  10^6$) the rotational dynamics exhibits no substantial revivals, indicating  that the rotational wavepacket has  largely decohered due to the interaction with the nuclear spin environment.  Thus the internal environment of the molecule serves as a natural quantum simulator for controlled decoherence, in this case by external fields.

 \begin{figure}[t]
 \centering
 \includegraphics[width=0.83\columnwidth, trim = 20 15 0 -10]{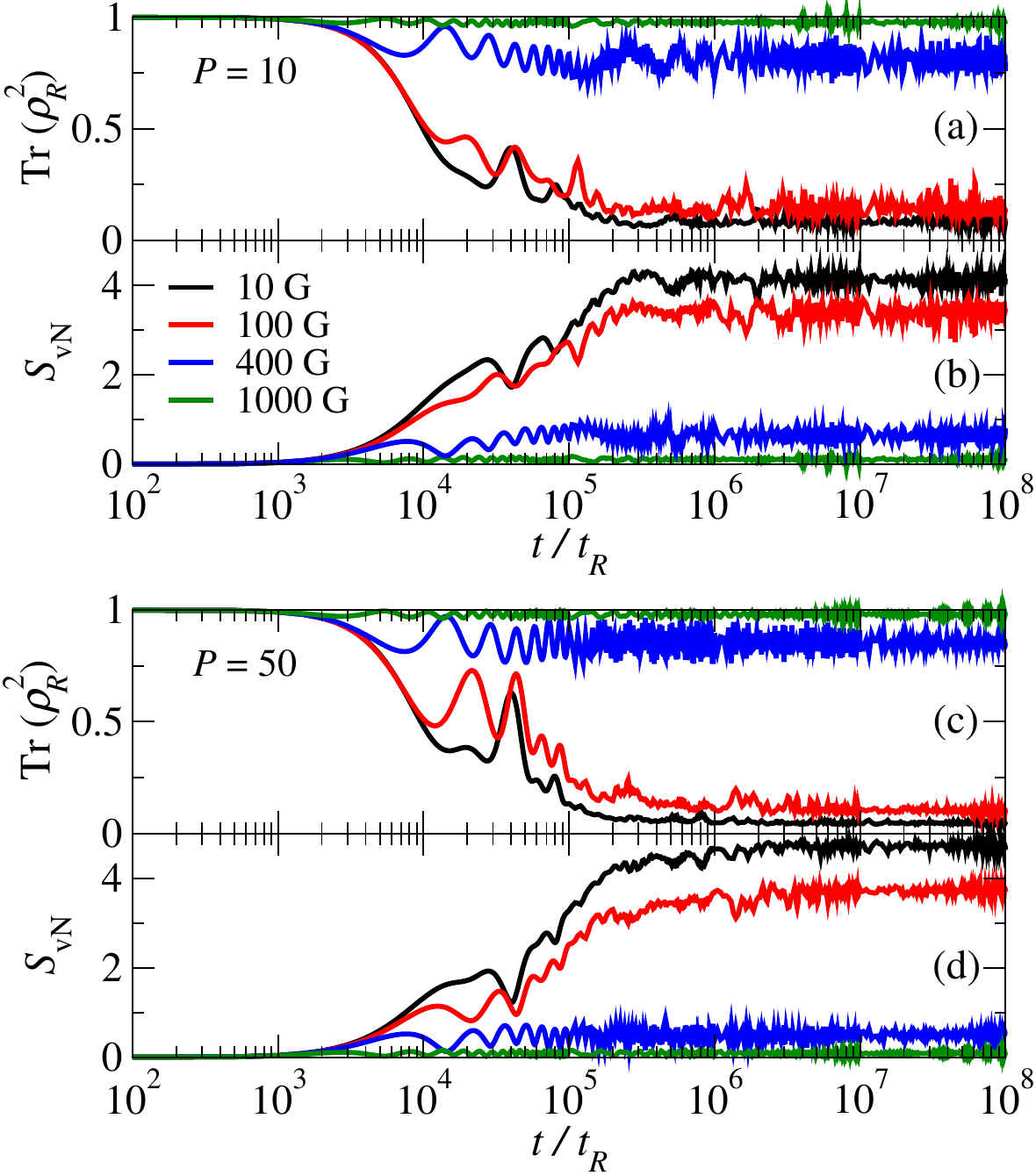}
 \caption{Time evolution of (a) the purity and (b) von Neumann entanglement entropy of the reduced density matrix $\hat{\rho}_R(t)$ for the rotationally ``warm'' KRb wavepacket created by a laser pulse with $P=10$. (c)-(d): Same as (a)-(b) but for the rotationally ``hot'' wavepacket with $P=50$. }
 \label{fig3}
\end{figure}

To further explore the decoherence dynamics, we plot the purity of the reduced density matrix, $\mathcal{P}(t)=\text{Tr}(\hat{\rho}^2_R)$ in Figs. 3(a) and 3(d). The rotational subsystem's quantum state remains nearly pure at early times ($\mathcal{P}\simeq 1$) due to the weak system-environment interactions (see above). At low $B$ fields, the purity exhibits a sharp decline at  $t/t_R\simeq 10^4$   as the NEQ interaction begins to  couple mechanical rotation to the nuclear spin environment. After a few revivals,  the purity approaches a quasi-steady state, in which small oscillations are observed superimposed on a constant background low purity state.  Remarkably, the qualitative features of purity dynamics are extremely sensitive to the external magnetic field: As shown in Fig. 3(a), at $B=1000$~G, the purity never deviates from unity over the entire course of dynamical evolution. The results at $B=400$~G exemplify the transition to the high-field regime, where rotational decoherence is strongly suppressed. 
 The suppression occurs because in the high $B$-field limit, the NEQ interaction becomes small compared to the nuclear spin Zeeman interaction. As a result, the molecular eigenstates acquire a direct-product structure, $|i\rangle\simeq |NM_N\rangle |M_{I_1}M_{I_2}\rangle$, 
  rendering the system-environment dynamics separable.
 
To gain more insight into rotational thermalization,
we plot in 
Figs.~3(b) and (d) the time evolution of the von Neumann entanglement entropy $S_\text{vN}(t)=-\sum_i \rho_i(t)\log_2 \rho_i(t) $, where $\rho_i$ are the eigenvalues of $\hat{\rho}_R$.
At low magnetic fields, the combined system-environment state evolves from non-entangled ($S_\text{vN}=0$) to highly entangled ($S_\text{vN}\simeq 4$).  This dynamics correlates nearly exactly with that of the purity loss in Figs.~3(a) and (c), which shows that the thermalization occurs as a result of the build-up of entanglement between the rotational subsystem and the nuclear spin environment, as previously observed in quantum many-body systems \cite{LewisSwan:19}.

In summary, we have introduced a new system-environment model (CRM) to describe rotational dynamics in molecules,
 which features
a rotational subsystem coupled by hyperfine interactions to the nuclear spin bath formed by the nuclei within the molecule. 
The CRM is similar to the paradigmatic CSM \cite{Prokofev:00}, but features important differences due to the infinite-dimensional Hilbert space of the rotational subsystem. 
The proposed analogy between the CRM and CSM suggests that molecular rotational dynamics can be simulated using the powerful techniques already developed for the CSM (such as cluster expansions~\cite{Witzel:08,Witzel:12}). 
This opens up novel opportunities for exploring these  dynamics in large polyatomic molecules, where ED calculations would be impractical, such as $^{13}${C}$_{60}$~\cite{Jex:00}, whose Hilbert spaces may offer unique advantages for quantum science applications~\cite{Albert:24}.
Ultracold molecules provide a native platform for quantum simulation of the CRM, opening up new directions of research in quantum information dynamics with ultracold molecular gases.

We thank Dr. X. Xing for his assistance in preparing Fig.~1. 
This work was performed in part with support from the NSF under CAREER grant PHY-2045681 (TVT), grant PHY-22010566 (LDC), and grant PHY-2309135 to the Kavli Institute for Theoretical Physics (KITP) (LDC).

\end{document}